\documentclass[twocolumn,twoside]{revtex4}
\usepackage{graphicx}
\usepackage{fancyhdr}
\begin{document}

\title{\Large Indirect searches for dark matter annihilations toward\\ dwarf spheroidal galaxies with VERITAS}

%

\author{M. Vivier}
\affiliation{Department of Physics \& Astronomy and the Bartol Research Institute, University of Delaware, Newark DE 19716}
\author{for the VERITAS collaboration}
\affiliation{http://veritas.sao.arizona.edu/}

\begin{abstract}
In the cosmological paradigm, cold Dark Matter (DM) dominates the mass content of the Universe and is present at every scale. Candidates for DM include many extensions of the standard model, with a weakly interacting massive particle (WIMP) in the mass range from 50 GeV to greater than 10 TeV. The self-annihilation of WIMPs in astrophysical regions of high DM density can produce secondary particles, including very high energy (VHE) $\mathrm{\gamma}$-rays, with energies up to the dark matter particle mass. The VERITAS array of Cherenkov telescopes, designed for the detection of VHE $\mathrm{\gamma}$-rays in the 100 GeV-10 TeV energy range, is an appropriate instrument for the detection of DM and is complementary to Fermi-LAT. Dwarf spheroidal galaxies (dSphs) of the Local Group are potentially the best targets to search for the annihilation signature of DM due to their proximity and large DM content. We report on the latest VERITAS observations of dSphs and discuss the results in the framework of WIMP models.
\end{abstract}

\maketitle

\thispagestyle{fancy}
\section{Introduction}
\noindent The compelling evidence for the presence of DM in the different structures of the Universe \cite{DM} has motivated numerous efforts to search for DM by means of astrophysical observations. If DM is made of WIMPs annihilating into standard model particles, indirect searches for DM annihilations with VHE $\mathrm{\gamma}$-rays provide a very promising way to constrain the nature of the DM particle: $\mathrm{\gamma}$-rays are free of any propagation effects on short distances ($\mathrm{\leq}$ 1\,Mpc) and DM particle annihilation is predicted to give a unique $\mathrm{\gamma}$-ray spectrum. Such searches are often conducted using pointed observations toward nearby DM overdensities, because the annihilation rate is proportional to the squared DM density. Popular targets include the Galactic Center \cite{GCBuckley, GCWhipple,GCDM1,GCDM2}, satellite galaxies of the Milky-Way (MW)  \cite{dSphDM, dSph1,dSph2,dSph3,dSph4,dSph5,MAGICSegue1}, globular clusters \cite{GloC1,GloC2} and clusters of galaxies \cite{Cluster}.\\
The dSphs of the Local Group best meet the criteria for a clear and unambiguous detection of DM. They are gravitationally bound objects and contain up to $\cal{O}$($\mathrm{10^{3}}$) more mass in DM than in visible matter. As opposed to the Galactic Center, they are environments with a favorably low astrophysical $\mathrm{\gamma}$-ray background. Neither astrophysical $\mathrm{\gamma}$-ray sources (supernova remnants, pulsar wind nebulae, etc) nor gas acting as target material for cosmic rays have been observed in these systems \cite{dSph}. Furthermore, their relative proximity and high galactic latitude make them the best astrophysical targets for high signal-to-noise detection.\\
This paper gives a status of the dSph observational program carried out by the ground-based VHE $\mathrm{\gamma}$-ray observatory VERITAS. Section \ref{Sec2} presents the instrument and summarizes the VERITAS dSph observations and data analysis. Section \ref{Sec3} interprets the results in terms of constraints on DM models, focusing on models which include a Sommerfeld enhancement. Finally, section \ref{Sec4} is devoted to the conclusion.
\section{VERITAS observation and analysis}\label{Sec2}
\noindent VERITAS is an array of four 12-meter imaging atmospheric Cherenkov telescopes (IACTs) located at the base camp of the F. L. Whipple Observatory in southern Arizona. Each VERITAS telescope consists of a large optical reflector which focuses the Cherenkov light emitted by particle air showers onto a camera of 499 photomultiplier tubes. The large effective area ($\mathrm{\sim 10^{5}\,m^{2}}$), in conjunction with the stereoscopic imaging of air showers, enables VERITAS to be sensitive over a wide range of energies (from 100 GeV to 30 TeV) with an energy and angular resolution of 15-20\% and $\mathrm{0.1^{\circ}}$ per reconstructed $\gamma$-ray, respectively. VERITAS is able to detect a point source with 1\% of the Crab Nebula flux at a statistical significance of 5 standard deviations above background ($\mathrm{5\,\sigma}$) in approximately 25 hours of observations. For further details about VERITAS, see, e.g., \cite{VERITAS}.
\begin{table*}[!ht]
\begin{center}
\begin{tabular}{|c|c|c|c|c|c|}
\hline \textbf{dSph target} & \textbf{Draco} & \textbf{Ursa Minor} & \textbf{Bootes 1} & \textbf{Willman 1} & \textbf{Segue 1}\\
\hline
\hline 
Exposure [hours] & 18.4 & 18.9 & 14.3 & 13.7 & 51.5\\
\hline 
Zenith angle [deg]  & 26-51 & 35-46 & 17-29 & 19-28 & 16-39 \\
\hline 
Excess [counts] & -28.4 & -30.4 & 28.5 & -1.45 & 30.4\\
\hline 
Significance [$\mathrm{\sigma}$] & -1.5 & -1.8 & 1.3 & -0.1 & 0.9\\
\hline 
95\% CL upper limit [counts] & 18.8 & 15.6 & 72.0 & 36.7 & 135.9 \\
\hline 
$\mathrm{\Phi^{95\%\,CL}(E\geq 300\,GeV)}$ [$\mathrm{cm^{-2}\,s^{-1}}$] & $\mathrm{0.5\times 10^{-12}}$ & $\mathrm{0.4\times 10^{-12}}$ & $\mathrm{2.2\times 10^{-12}}$ & $\mathrm{1.2\times 10^{-12}}$ & $\mathrm{0.8\times 10^{-12}}$\\
\hline
\end{tabular}
\caption{The VERITAS dSph observations and corresponding analysis results. The significance is calculated according to the method of Li \& Ma \cite{LiMa}. The 95\% CL ULs on the number of $\mathrm{\gamma}$-rays in the ON-source region is derived using the Rolke prescription \cite{RolkeMethod}. The ULs on the integrated flux have been computed assuming a power law of index -2.6.}\label{Tab1}
\end{center}
\end{table*}\\
Since 2007, the VERITAS observatory has started an active observational program on a dSph sample representing the most promising targets visible in the northern hemisphere. Table \ref{Tab1} presents the current observations carried out by VERITAS on five well-studied and closeby dSphs, namely the Draco, Ursa Minor, Bootes 1, Willman 1 and Segue 1 galaxies. The so-called wobble pointing mode, where the pointing position is shifted by $\mathrm{\pm}$ 0.5 deg from the target position, has been used during these observations. The wobble pointing strategy allows for simultaneous background  estimation and source observation, thus reducing the systematic uncertainties in the background determination. Data reduction follows the methods described in \cite{VERITASdataanalysis}. After calibration of the data, the $\mathrm{\gamma}$-rays are selected by calculating the Hillas parameters of the recorded camera images. The selection cuts were optimized for the detection of a 5\% Crab Nebula-like source. After the $\mathrm{\gamma}$-ray selection, the residual hadronic background in a circular region of 0.12 deg radius centered on the target position (called the ON-source region) is estimated using the reflected background model. The analysis of the data did not reveal any significant excess over the estimated background in any of these observations. Table \ref{Tab1} displays the measured $\mathrm{\gamma}$-ray excesses for each of the dSph dataset, along with the corresponding significances, the resulting upper limits (ULs) on the number of $\mathrm{\gamma}$-rays in the ON source region and on the integrated flux above 300 GeV. The 95\% confidence level (CL) integral flux ULs lie in the range 0.4-1.2\% of the Crab Nebula flux.
\section{Constraints on Dark Matter models}\label{Sec3}
\subsection{Classical constraints}
\begin{figure}[!bt]
\includegraphics[width=3in]{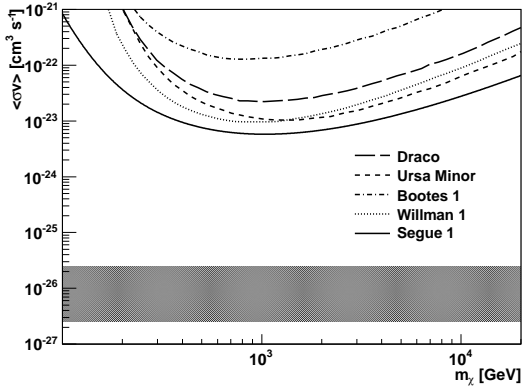}
\caption{95\% CL ULs on $\mathrm{\langle\sigma v\rangle}$ as a function of the DM particle mass for the five dSphs observed with VERITAS . The limits are computed using a composite DM annihilation spectrum (10\% $\mathrm{\tau^{+}\tau^{-}}$ + 90\% $\mathrm{b\bar{b}}$), except for the case of Segue 1 in which a pure $\mathrm{W^{+}W^{-}}$ channel was considered. The grey band area represents a range of generic values for the annihilation cross-section for thermally produced DM.}\label{Fig1}
\end{figure}

\begin{table*}[!ht]
\begin{center}
\begin{tabular}{|c|c|c|c|c|c|}
\hline \textbf{dSph target} & \textbf{Draco} & \textbf{Ursa Minor} & \textbf{Bootes 1} & \textbf{Willman 1} & \textbf{Segue 1}\\
\hline
\hline 
Profile modeling & NFW & NFW & NFW & NFW & Einasto\\
\hline 
Distance [kpc]  & 80 & 66 & 62 & 38 & 23 \\
\hline 
$\mathrm{\bar{J}(\theta\leq0.12^{\circ})}$ [$\mathrm{GeV^{2}\,cm^{-5}\,sr}$] & $\mathrm{1.5\times10^{18}}$ & $\mathrm{2.7\times10^{18}}$ & $\mathrm{1.1\times10^{18}}$ & $\mathrm{8.4\times10^{18}}$ & $\mathrm{7.7\times10^{18}}$\\
\hline 
\end{tabular}
\caption{Profile modeling, distance and astrophysical factors of the five dSphs observed with VERITAS.}\label{Tab2}
\end{center}
\end{table*}
\noindent The absence of signal in any of these observations can be used to derive constraints on various dark matter models. The $\mathrm{\gamma}$-ray flux from the annihilations of DM particles, of mass $\mathrm{m_{DM}}$, in a spherical DM halo is given by a particle physics term times an astrophysics term:
\begin{equation}
\mathrm{\frac{d\Phi_{\gamma}}{dE}(\Delta\Omega,E)=\frac{\langle\sigma v \rangle}{8\,\pi\,m_{DM}^{2}}\,\frac{dN_{\gamma}}{dE}\times\bar{J}(\Delta\Omega)}.
\end{equation}
The particle physics term contains all the information about the DM particle: its mass $\mathrm{m_{DM}}$, its total velocity-weighted annihilation cross-section $\mathrm{\langle\sigma v \rangle}$ and the differential $\mathrm{\gamma}$-ray spectrum from all final states weighted by their corresponding branching ratios, $\mathrm{dN_{\gamma}/dE}$. The astrophysical factor $\mathrm{\bar{J}(\Delta\Omega)}$ is the square of the DM density integrated along the line of sight (s) and over the solid angle $\mathrm{\Delta\Omega}$:
\begin{equation}
\mathrm{\bar{J}(\Delta \Omega) = \int_{\Delta \Omega} d\Omega \int_{los}\rho_{\chi}^{2}(r[s])ds.}\label{fAP}
\end{equation}
The solid angle is given here by the size of the signal search region defined previously in the analysis, i.e. $\mathrm{\theta}$ $\mathrm{\leq}$ 0.12 deg. The estimate of the astrophysical factor requires a modeling of the dSph DM distribution. Each dSph DM distribution has been modeled with a Navarro, Frenk \& White (NFW) profile \cite{NFW}, except Segue 1 for which the DM distribution has been modeled with an Einasto profile \cite{Einasto}. The parameters of a dSph profile are constrained using its star kinematics and have each been taken from the literature \cite{dSph1,Segue1HL2}. The resulting astrophysical factors of Draco, Ursa Minor, Bootes 1, Willman 1 and Segue 1 are presented in Table \ref{Tab2}, along with their distance and profile modeling.\\
Once the astrophysical factors have been estimated, the ULs in the number of $\gamma$-ray candidates for each dSph can be translated into ULs on the total annihilation cross-section:
\begin{equation}
\mathrm{\langle\sigma v\rangle^{95\%\,CL}= \frac{8\pi}{\bar{J}(\Delta\Omega)}\times} \\
\mathrm{\frac{N_{\gamma}^{95\% CL}m_{DM}^{2}}{T_{obs}\,\int_0^{m_{DM}}{\cal{A}}_{eff}(E)\frac{dN_{\gamma}}{dE}dE}},
\end{equation}
where $\mathrm{T_{obs}}$ is the total observation time and $\mathrm{{\cal{A}}_{eff}(E)}$ is the effective area as a function of energy, zenith angle and the offset of the source from the pointing position.\\
Figure \ref{Fig1} shows the VERITAS 95\% CL ULs on the velocity-weitghted annihilation cross-section $\mathrm{\langle \sigma v \rangle}$ as a function of the DM particle mass for each of the five observed dSph listed in Table \ref{Tab1}. The Draco, Ursa Minor, Bootes 1 and Willman 1 limits have been obtained using a composite DM spectrum with a 90 \%  $\mathrm{b\bar{b}}$ branching ratio and a 10\% $\mathrm{\tau^{+}\tau^{-}}$ branching ratio \cite{dSph1}. The Segue 1 limits have been obtained assuming a pure $\mathrm{W^{+}W^{-}}$ DM annihilation spectrum, which is similar to the composite $\mathrm{b\bar{b}}$ - $\mathrm{\tau^{+}\tau^{-}}$ spectrum. The limits are in the range $\mathrm{10^{-21}-10^{-23}\,cm^{3}\,s^{-1}}$, depending on the considered dSph. The best limits are obtained for the Segue 1 dSph, which has the deepest exposure and one of the highest astrophysical factor among the five dSphs (see table \ref{Tab2}). The 95\% CL upper limit on $\mathrm{\langle\sigma v\rangle}$ is at the level of $\mathrm{8\times10^{-24}\,cm^{3}\,s^{-1}}$ around 1 TeV and is among the best reported so far with dSph VHE observations.
\subsection{Constraints on models with a Sommerfeld enhancement}
\begin{figure*}[!ht]
\centerline{\includegraphics[width=3in]{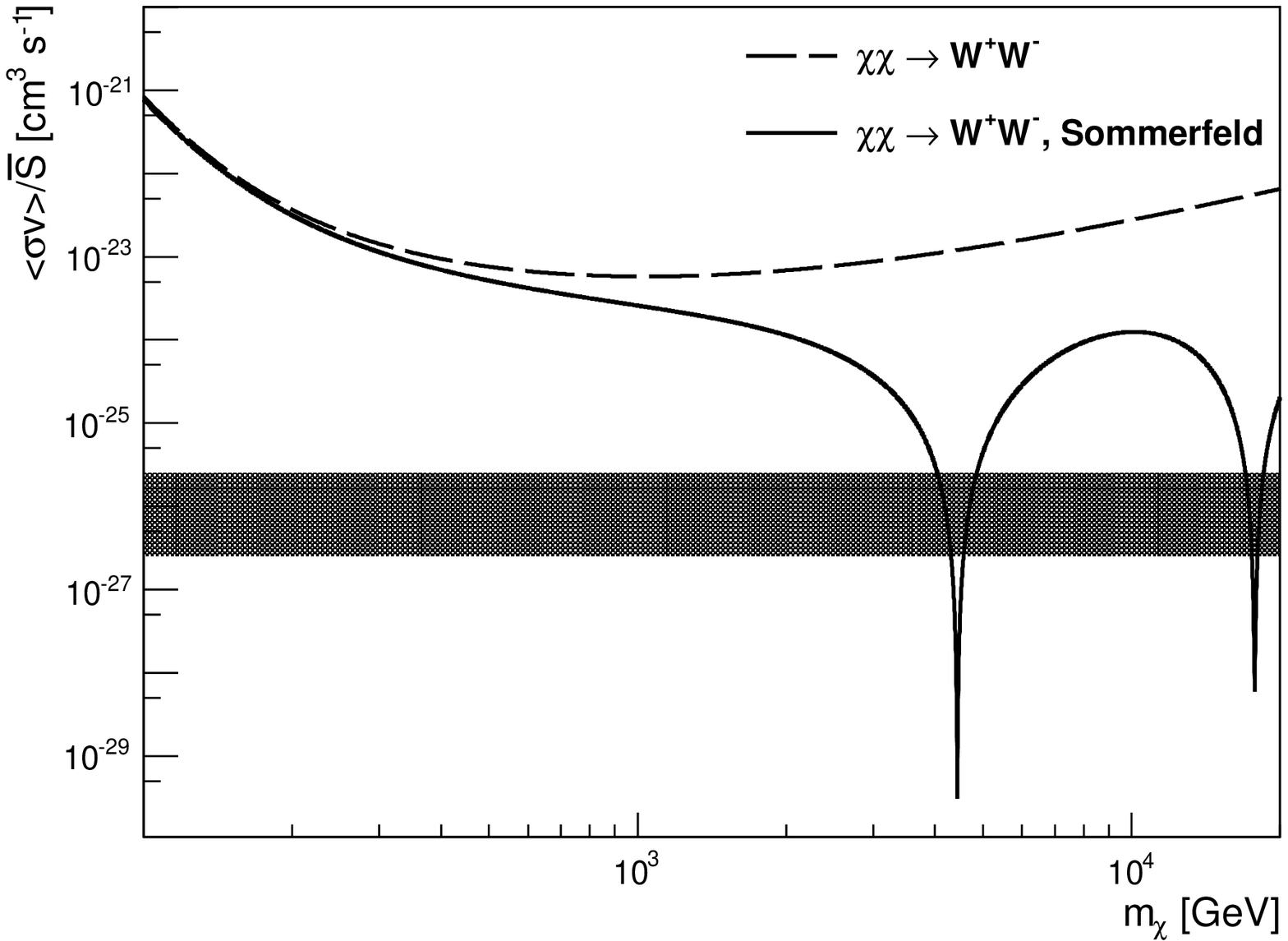}
              \hfil
              \includegraphics[width=3in]{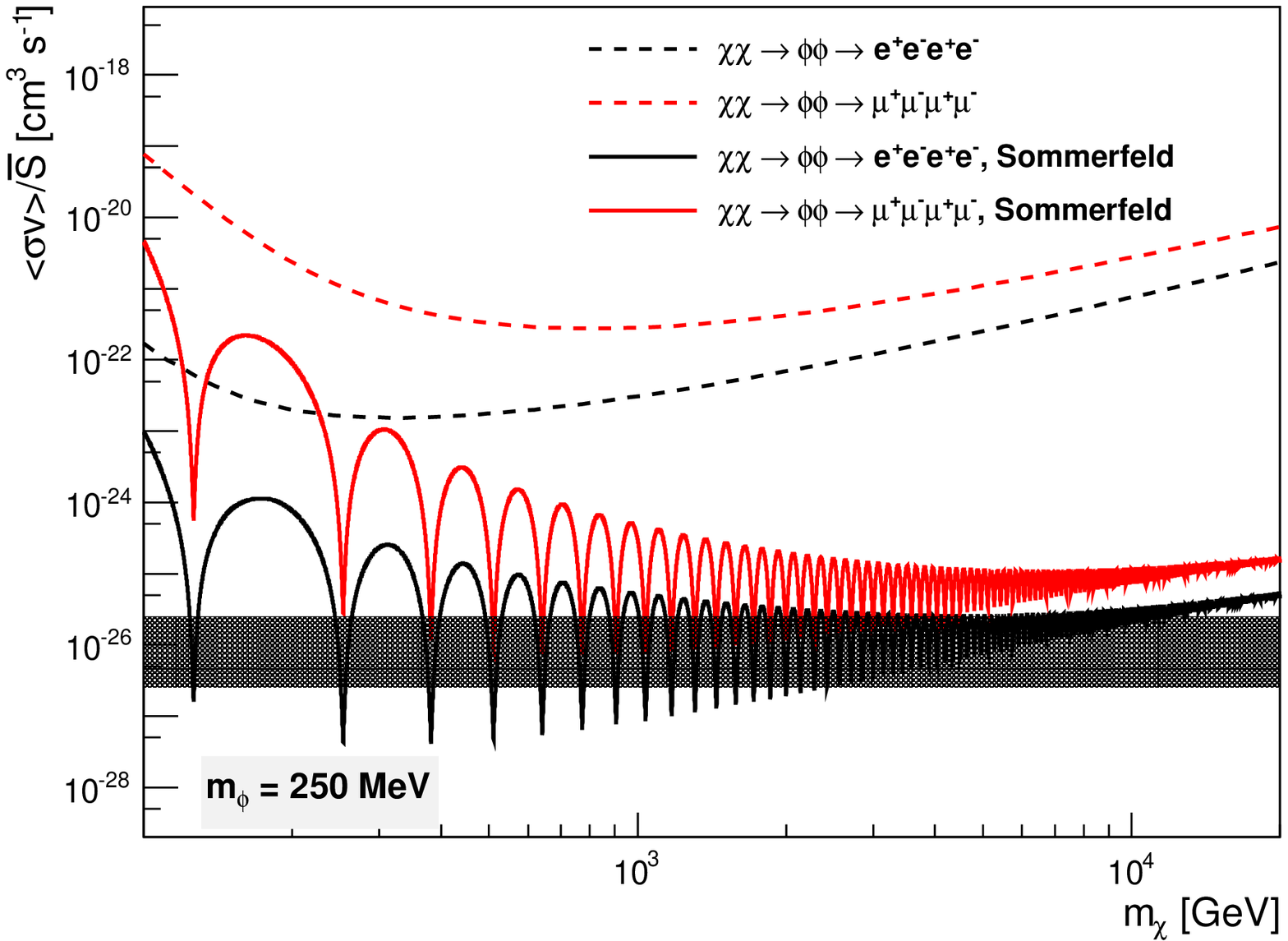}
             }
\caption{95\% CL exclusion curves from the VERITAS observations of Segue 1 on $\mathrm{\langle \sigma v \rangle}$/$\mathrm{\bar{S}}$ as a function of the dark matter particle mass, in the framework of two models with a Sommerfeld enhancement. The expected Sommerfeld enhancement $\mathrm{\bar{S}}$ applied to the particular case of Segue 1 has been computed assuming a Maxwellian dark matter relative velocity distribution \cite{Segue1HL3}. The grey band area represents a range of generic values for the annihilation cross-section in the case of thermally produced dark matter. Left: model I with wino-like neutralino dark matter annihilating to a pair of $\mathrm{W^{+}W^{-}}$ bosons. Right: model II with a 250 MeV scalar particle decaying into either $\mathrm{e^{+}e^{-}}$ or $\mathrm{\mu^{+}\mu^{-}}$.} \label{Fig2}
\end{figure*}

\noindent In this section, we consider DM models where the annihilation cross-section can be boosted with the Sommerfeld enhancement effect. We use the Segue 1 data to constrain such models. The Sommerfeld enhancement is a particle physics effect that was suggested to explain the cosmic ray lepton anomalies measured by ATIC and PAMELA \cite{ATIC,PAMELApositronfraction}. It arises when the two DM particles interact through an attractive potential, mediated by the exchange of a massive boson \cite{SommerfeldW}. The annihilation cross-section enhancement is particularly effective when the DM particle relative velocity is very small. Depending on the mass $\mathrm{m_{\phi}}$ and the coupling $\mathrm{\alpha}$ of the exchanged boson, the Sommerfeld enhancement can exhibit a series of resonances for specific values of the DM particle mass, giving very large boost factors up to $\mathrm{10^{6}}$. In such scenarios, the annihilation cross-section is given by:
\begin{equation}
\mathrm{\langle\sigma v\rangle=\bar{S}\times (\sigma v)_0},
\end{equation}
where $\mathrm{\bar{S}}$ is the Sommerfeld boost factor averaged over the DM particle velocity distribution and depends on the DM particle mass. $\mathrm{(\sigma v)_0}$ is the thermal WIMP annihilation cross-section before freeze-out. The Sommerfeld enhancement is of particular interest for cold DM halos like dSphs, where the mean star velocity dispersion can reach a few $\mathrm{km\,s^{-1}}$.\\
We focus here on two models comprising a Sommerfeld enhancement to the annihilation cross-section. The first model (hereafter model I, \cite{SommerfeldW}) assumes that the dark matter particle is a wino-like neutralino $\mathrm{\chi^{0}}$ (arising in SUSY extensions of the standard model), which annihilates through the exchange of a $\mathrm{Z^{0}}$ boson ($\mathrm{m_{Z^{0}} \sim 90\,GeV}$, $\mathrm{\alpha \sim 1/30}$), thus leading to a Sommerfeld enhancement. The second model (hereafter model II) introduces a new force in the dark sector \cite{AH}, carried by a light scalar field $\mathrm{\phi}$ predominantly decaying into leptons and with a mass $\mathrm{{\cal{O}}(1\,GeV)}$. In such models, dark matter annihilates to a pair of $\mathrm{\phi}$ scalar particles, with an annihilation cross-section boosted by the Sommerfeld enhancement. The coupling $\mathrm{\alpha}$ of the light scalar particle $\mathrm{\phi}$ to the dark matter particle is determined assuming that $\mathrm{\chi\chi \rightarrow \phi\phi}$ is the only channel that regulates the dark matter density before freeze-out \cite{Hulthen2}. Figure \ref{Fig2} shows the VERITAS constraints on each of these models, derived with the observations of Segue 1. The left panel of Figure \ref{Fig2} shows the constraints on model I. The Sommerfeld enhancement exhibits two resonances in the considered dark matter particle mass range, for $\mathrm{m_{\chi} \simeq 4.5\,TeV}$ and $\mathrm{m_{\chi} \simeq 17\,TeV}$, respectively. VERITAS excludes these resonances, which boost the annihilation cross-section far beyond the canonical $\mathrm{\langle \sigma v \rangle \sim 3\times10^{-26}\,cm^{3}\,s^{-1}}$. The right panel of Figure \ref{Fig2} shows the VERITAS constraints on model II, for a scalar particle with mass $\mathrm{m_{\phi} = 250\,MeV}$. Two channels in which the scalar particle decays either to $\mathrm{e^{+}e^{-}}$ or $\mathrm{\mu^{+}\mu^{-}}$ have been considered. VERITAS observations start to disfavor such models, especially for the $\mathrm{e^{+}e^{-}e^{+}e^{-}}$ channel where some of the resonances are beyond $\mathrm{\langle \sigma v \rangle \sim 3\times10^{-26}\,cm^{3}\,s^{-1}}$. This result holds for $\mathrm{\phi}$ particle masses up to a few GeV.
\section{Summary and conclusion}\label{Sec4}
\noindent Since its commissioning phase, the VERITAS observatory has been actively observing the best dSph targets visible from the northern hemisphere, with data currently taken toward the Draco, Ursa Minor, Bootes 1, Willman 1 and Segue 1 galaxies. The non-detection of a VHE $\mathrm{\gamma}$-ray signal in any of these observations has lead to limits on the velocity-weighted annihilation cross-section $\mathrm{\langle\sigma v \rangle}$ of the order of $\mathrm{10^{-21}-10^{-23}\,cm^{3}\,s^{-1}}$. The Segue 1 data have provided the best limits on $\mathrm{\langle\sigma v \rangle}$ reported so far with any dSph observations, though these limits are at least two orders of magnitude away from the canonical $\mathrm{\langle \sigma v \rangle \sim 3\times10^{-26}\,cm^{3}\,s^{-1}}$. The Segue 1 data have also been used to constrain models with a Sommerfeld enhancement, disfavoring them as being a good interpretation of the cosmic ray lepton excesses measured by ATIC and PAMELA. 
\begin{acknowledgments}
\noindent This research was supported by grants from the U.S. Department of Energy, the U.S. National Science Foundation and the Smithsonian Institution, by NSERC in Canada, by Science Foundation Ireland and by the Science and Technology Facilities Council in the UK. We acknowledge the work of the technical support staff at the Fred Lawrence Whipple Observatory.
\end{acknowledgments}

\end{document}